\documentclass[12pt,twoside]{article}
\usepackage[mathscr]{eucal}
\usepackage{amsmath,amsfonts,amssymb,amsthm,empheq}
\usepackage{times}
\usepackage{pdfsync}
\usepackage{cite}
\usepackage{url}
\usepackage{hyperref}
\usepackage{tensor}
\usepackage{color}
\allowdisplaybreaks[4]

\advance\textheight\topskip
\def\nn{\nonumber}

\def\bP{\bar{P}}

\def\p{\partial}
\def\bz{\bar z}

\allowdisplaybreaks[1]

\def\bm{\bar m}
\newcommand{\bea}{\begin{eqnarray}}
\newcommand{\eea}{\end{eqnarray}}
\newcommand{\be}{\begin{equation}}
\newcommand{\ee}{\end{equation}}
\newcommand{\bs}{\begin{split}}
\newcommand{\es}{\end{split}}

\newcommand{\ffrac}[2]{\raisebox{.5pt}%
  {\footnotesize$\displaystyle\frac{#1}{#2}$}\kern1pt}

\newcommand{\n}{\nabla}

\def\cL{\mathcal{L}}

\def\cO{\mathcal{O}}

\numberwithin{equation}{section} \makeatletter
\@addtoreset{equation}{section}

\DeclareFontFamily{OT1}{rsfs}{} \DeclareFontShape{OT1}{rsfs}{m}{n}{
<-7> rsfs5 <7-10> rsfs7 <10-> rsfs10}{}
\DeclareMathAlphabet{\mycal}{OT1}{rsfs}{m}{n}

\newcommand*\xbar[1]{%
  \hbox{%
    \vbox{%
      \hrule height 0.5pt 
      \kern0.3ex
      \hbox{%
        \kern-0.0em
        \ensuremath{#1}%
        \kern-0.0em
      }%
    }%
  }%
}
\newcommand\email[1]{\thanks{\href{mailto:#1}{\nolinkurl{#1}}}}

\begin{document}

\title{Einstein-Maxwell-dilaton theory in Newman-Penrose formalism}

\author{Wen-Di Tan}

\date{}

\def\mytitle{Einstein-Maxwell-dilaton theory in Newman-Penrose formalism}

\pagestyle{myheadings} \markboth{\textsc{\small W.~Tan}}
{\textsc{\small Einstein-Maxwell-dilaton theory in Newman-Penrose formalism}}

\addtolength{\headsep}{4pt}

\begin{centering}

  \vspace{1cm}

  \textbf{\Large{\mytitle}}

  \vspace{1.5cm}

  {\large Wen-Di Tan}

\vspace{.5cm}

\vspace{.5cm}
\begin{minipage}{.9\textwidth}\small \it  \begin{center}
     Center for Joint Quantum Studies and Department of Physics,\\
     School of Science, Tianjin University, 135 Yaguan Road, Tianjin 300350, China
 \end{center}
\end{minipage}

\end{centering}


\vspace{1cm}

\begin{center}
\begin{minipage}{.9\textwidth}

  \textsc{Abstract}. In this paper, we study the four dimensional Einstein-Maxwell-dilaton (EMD) theories in the Newman-Penrose (NP) formalism. We adapt the equations of motion into the NP formalism, and obtain the solution space that is asymptotic to the flat space-time. We then investigate the gravitational and electromagnetic memory effects.  We find that the dilaton does not contribute to the displacement nor the kick memory effects, but it does contribute to the time-delayed memory effect.
 \end{minipage}
\end{center}


\thispagestyle{empty}


\section{Introduction}
\label{sec:introduction}

In 1960s, to understand the gravitational radiation in full Einstein theory, Bondi and his collaborators established an elegant framework for axisymmetric isolated systems and demonstrated that gravitational waves exist in the full Einstein theory rather than an artifact of linearization\cite{BMS}. In this framework, they chose a suitable coordinates system and expanded the metric fields in inverse powers of the radial coordinate $r$. Imposing the proper boundary conditions, the equations of motion can be solved order by order in $1/r$ expansions. In this framework, the gravitational radiation is characterized by the news functions and the mass of the system decreases whenever the news function exists. Shortly, this framework was extended to asymptotically flat space-times by Sachs\cite{Sachs}. Meanwhile, Newman and Penrose\cite{NP} developed a new approach to understand gravitational radiation by means of a tetrad or spinor formalism. They derived a compact set of first order differential equations involving linear combinations of the equations for the Riemann tensor, expressed in the Ricci rotation coefficients or the spinor affine connection. These equations are equivalent to the empty space Einstein equations. From these equations, one can investigate the asymptotic behavior of the fields systematically, under the condition that the space-time should approach to flatness at infinity. The asymptotic flatness condition is imposed on the Riemann tensor rather than the metric.

This formalism is motivated by the strong belief that the essential element of a space-time is its light-cone structure and it is the most effective way for grasping the inherent symmetries of the space-times such as the black-hole solutions of general relativities. In this formalism, the geometrical property of the space-times is more transparent and it is the most satisfactory way to study the fermion-coupled theories. The asymptotically flat solutions of the empty Newman-Penrose equations were later derived by Newman and Unti\cite{NU}, and the News functions and the mass-loss formula were successfully recovered. When matter fields are coupled, one expects that the matter equations of motion are also adaptable into the NP formalism. This was indeed done for Einstein-Maxwell gravity \cite{BK,ENP}. However other theories are less studied in NP formalism.

In this paper, we study the four dimensional Einstein-Maxwell-dilaton (EMD) theories in Newman-Penrose formalism. Including the Kaluza-Klein theory which arises from five dimensional Einstein gravity reduced on a circle, the four-dimensional EMD theories are a class of theories that can be embedded into various supergravities which originate from string theories or M-theory. In these theories, the matter sector includes the Maxwell field $A$ and a dilatonic scalar $\varphi$, both of which are massless and minimally coupled to gravity. The dilaton is non-minimally coupled to the Maxwell kinetic term in the form of an exponential function $e^{a\varphi}$ where $a$ is the dilaton coupling constant. In\cite{HLC}, the authors used Newman-Penrose formalism to analyze the perturbations of the Kerr-Newman dilatonic black hole background. However, EMD theories have not been fully studied in NP formalism elsewhere. In this paper, we adapt the equations of motion of EMD theories into NP formalism and obtain the solution space that approaches the flatness asymptotically. We then examine the mass-loss formula and the charge conservation. As a direct application, we also study the memory effects.

First reported by Zel'dovich and Polnarev\cite{ZP} in linearized gravity and further studied by Christodoulou in full Einstein gravity\cite{C}, gravitational memory effects are a large group of observational effects for gravitational radiation which is characterized by the change of the asymptotic shear $\Delta\sigma^{0}$\cite{F} (see also \cite{BG,BT,WW,T,LTLBC,N,YM} for relevant developments). The memory effects also exist in Maxwell theory named electromagnetic memory effects\cite{SUS,LBDG}. We should notice that the memory effects can be classified by the observational effects.

In the recent years, there have been renewed interests in memory effects. Strominger and Zhiboedov\cite{SZ} discovered an intriguing triangular relation of three ingredients: the BMS supertranslation symmetry, the leading soft graviton theorem and a displacement gravitational memory effect. This memory effect is a displacement of two parallel inertial detectors caused by the radiative energy flux and it turns out to be mathematically equivalent to Weinberg's soft graviton theorem\cite{W} by the Fourier or the inverse Fourier transformations.

Pasterski, Strominger and Zhiboedov\cite{PSZ} later discovered a spin memory effect which characterized by the relative time delay between the different orbiting light rays caused by the radiative angular momentum. This spin memory effect was shown to be mathematically equivalent to the subleading soft graviton theorem\cite{SC}. In \cite{MW}, the authors found another spin memory effect represented by the proper time delay of a free-falling massive particle constrained on a time-like $r=r_{0}$ hypersurface near the null infinity. In \cite{MT}, the authors considered the motion of a charged observer and investigated both the gravitational and electromagnetic memory effects in a unified manner in the Einstein-Maxwell theory. It is thus of great interest to study the memory effect in EMD theories to uncover the possible observational effects due to the non-minimal coupling between the scalar and the electromagnetic field. The fully understood memory effects in EMD theories may also help to understand the memory effects in string theories \cite{AES} and M-theory.

The paper is organized as follows. In section 2, we will give a brief introduction of NP formalism and derive the NP equations of the four-dimensional Einstein-Maxwell-dilaton theories. The asymptotically-flat solution space of these theories will be derived in section 3. We also examine the charge conservation and the mass-loss formula. In section 4, we will obtain the memory effects based on the investigation of \cite{MT} in EMD theories. We conclude the paper in section 5.

\section{Einstein-Maxwell-dilaton theory in the NP formalism}
\label{EM}
 The Newman-Penrose formalism is a special tetrad formalism with two real null basis vectors $e_{1}=l$, $e_{2}=n$, and two complex null basis vectors $e_{3}=m$, $e_{4}=\bm$. These basis vectors have the orthogonality relations
\begin{equation}
\label{eq1}
l\cdot m=l\cdot\bm=n\cdot m=n\cdot\bm=0,
\end{equation}
and are normalized as
\begin{equation}
\label{eq2}
l\cdot n=1,\quad m\cdot\bm=-1.
\end{equation}
The metric is obtained from the basis vectors as
\be
\label{eq3}
g_{\mu\nu}=\eta_{ab}(\mathbf{e}^{a}_{\mu})(\mathbf{e}^{b}_{\nu})=n_\mu l_\nu + l_\mu n_\nu - m_\mu {\bm}_\nu - m_\nu \bm_\mu.
\ee
where $\mathbf{e}^{a}_{\mu}$ represents the basis vector $l, n, m, \bm$, $\mu$ is the coordinate index, while $a$ is the tetrad index, $\eta_{ab}$ is the metric component under the tetrad form. The connection coefficients, called spin coefficients in the NP formalism with special Greek symbols (we will follow the convention of \cite{Ch}), are presented as follows
\be\begin{split}
&\kappa=\Gamma_{311}=l^{\nu}m^{\mu}\nabla_{\nu}l_{\mu},\;\;\;\;\;\;\pi=-\Gamma_{421}-l^{\nu}\bm^{\mu}\nabla_{\nu}n_{\mu},\cr
&\epsilon=\frac{1}{2}(\Gamma_{211}-\Gamma_{431})=\frac{1}{2}(l^{\nu}n^{\mu}\nabla_{\nu}l_{\mu}-l^{\nu}\bm^{\mu}\nabla_{\nu}m_{\mu}).\label{eq4}
\end{split}\ee
\be\begin{split}
&\tau=\Gamma_{312}=n^{\nu}m^{\mu}\nabla_{\nu}l_{\mu},\;\;\;\;\;\;\nu=-\Gamma_{422}=-n^{\nu}\bm^{\mu}\nabla_{\nu}n_{\mu},\cr
&\gamma=\frac{1}{2}(\Gamma_{212}-\Gamma_{432})=\frac{1}{2}(n^{\nu}n^{\mu}\nabla_{\nu}l_{\mu}-n^{\nu}\bm^{\mu}\nabla_{\nu}m_{\mu}).\label{eq5}
\end{split}\ee
\be\begin{split}
&\sigma=\Gamma_{313}=m^{\nu}m^{\mu}\nabla_{\nu}l_{\mu},\;\;\;\;\;\;\mu=-\Gamma_{423}=-m^{\nu}\bm^{\mu}\nabla_{\nu}n_{\mu},\cr
&\beta=\frac{1}{2}(\Gamma_{213}-\Gamma_{433})=\frac{1}{2}(m^{\nu}n^{\mu}\nabla_{\nu}l_{\mu}-m^{\nu}\bm^{\mu}\nabla_{\nu}m_{\mu}).\label{eq6}
\end{split}\ee
\be\begin{split}
&\rho=\Gamma_{314}=\bm^{\nu}m^{\mu}\nabla_{\nu}l_{\mu},\;\;\;\;\;\;\lambda=-\Gamma_{424}=-\bm^{\nu}\bm^{\mu}\nabla_{\nu}n_{\mu},\cr
&\alpha=\frac{1}{2}(\Gamma_{214}-\Gamma_{434})=\frac{1}{2}(\bm^{\nu}n^{\mu}\nabla_{\nu}l_{\mu}-\bm^{\nu}\bm^{\mu}\nabla_{\nu}m_{\mu}).\label{eq7}
\end{split}\ee
We use five complex scalars to represent ten independent components of the Weyl tensors
\begin{equation}
\label{eq8}
\Psi_{0}=-C_{1313},\quad\Psi_{1}=-C_{1213},\quad\Psi_{2}=-C_{1342},\quad\Psi_{3}=-C_{1242},\quad\Psi_{4}=-C_{2424}.
\end{equation}
Ricci tensors are defined by four real and three complex scalars as follows
\bea
&&\Phi_{00}=-\frac{1}{2}R_{11},\;\;\Phi_{22}=-\frac{1}{2}R_{22},\;\;\Phi_{02}=-\frac{1}{2}R_{33},\;\;\Phi_{20}=-\frac{1}{2}R_{44},\cr
&&\Phi_{11}=-\frac{1}{4}(R_{12}+R_{34}),\;\;\Phi_{01}=-\frac{1}{2}R_{13},\;\;\Phi_{12}=-\frac{1}{2}R_{23},\cr
&&\Lambda=\frac{1}{24}R=\frac{1}{12}(R_{12}-R_{34}),\;\;\Phi_{10}=-\frac{1}{2}R_{14},\;\;\Phi_{21}=-\frac{1}{2}R_{24},\label{eq9}
\eea
where $\Lambda$ is the cosmological constant. Considered as directional derivatives, the basis vectors are represented by special symbols:
\begin{equation}
\label{eq10}
D=l^{\mu}\partial_{\mu},\quad \Delta=n^{\mu}\partial_{\mu},\quad \delta=m^{\mu}\partial_{\mu}.
\end{equation}

The equations that describing NP formalism include three classes:

(1)The commutation relations of the basis vectors and the structure constants
\begin{equation}
[\mathbf{e}_{a}, \mathbf{e}_{b}]=(\Gamma_{cba}-\Gamma_{cab})\mathbf{e}^{c}=C^{c}_{\ ab}\mathbf{e}_{c},
\end{equation}
where $\mathbf{e}_{a}$ is the basis vector, and $C^{c}_{\ ab}$ is the structure constant. The general tensorial formalism does not consider these relations since the coordinate basis is commutative. An example is as follows
\begin{equation}
\begin{aligned}{}
[\Delta, D]&=[n,l]=[\mathbf{e}_{2},\mathbf{e}_{1}]=(\Gamma_{c12}-\Gamma_{c21})\mathbf{e}^{c}\\
&=-\Gamma_{121}\Delta+\Gamma_{212}D-(\Gamma_{312}-\Gamma_{321})\xbar\delta-(\Gamma_{412}-\Gamma_{421})\delta.
\end{aligned}
\end{equation}
Giving the spin coefficients their symbols, we get
\begin{equation}
\Delta D-D\Delta=(\gamma+\xbar\gamma)D+(\epsilon+\xbar\epsilon)\Delta-(\xbar\tau+\pi)\delta-(\tau+\xbar\pi)\xbar\delta.
\end{equation}

(2)The Ricci identities, similar to using the coordinate basis to calculate the Riemann curvature tensor in the general tensorial formalism. i.e.
\begin{equation}
\begin{aligned}{}
-\Psi_{0}&=C_{1313}=R_{1313}=\Gamma_{133,1}-\Gamma_{131,3}\\
&+\Gamma_{133}(\Gamma_{121}+\Gamma_{431}-\Gamma_{413}+\Gamma_{431}+\Gamma_{134})\\
&-\Gamma_{131}(\Gamma_{433}+\Gamma_{123}-\Gamma_{213}+\Gamma_{231}+\Gamma_{132}).
\end{aligned}
\end{equation}
Substituting for the directional derivatives and the spin coefficients their designated symbols, we obtain
\begin{equation}
D\sigma-\delta\kappa=\sigma(3\epsilon-\xbar\epsilon+\rho+\xbar\rho)+\kappa(\xbar\pi-\tau-3\beta-\xbar\alpha)+\Psi_{0}.
\end{equation}

(3)The Bianchi identities. It is similar to the Bianchi identities in the tensorial form, i.e.
\begin{equation}
R_{1313|4}+R_{1334|1}+R_{1341|3}=0,
\end{equation}
where ``$\vert$'' represents the covariant derivative in tetrad form. It can be rewritten in the following form
\begin{equation}
-\xbar\delta\Psi_{0}+D\Psi_{1}+(4\alpha-\pi)\Psi_{0}-2(2\rho++\epsilon)\Psi_{1}+3\kappa\Psi_{2}+[Ricci]=0.
\end{equation}
Here
\begin{equation}
\begin{aligned}{}
[Ricci]&=-D\Phi_{01}+\delta\Phi_{00}+2(\epsilon+\xbar\rho)\Phi_{01}+2\sigma\Phi_{10}-2\kappa\Phi_{11}-\xbar\kappa\Phi_{02}\\
&+(\xbar\pi-2\xbar\alpha-2\beta)\Phi_{00}.
\end{aligned}
\end{equation}

As for Maxwell theory, in NP formalism the antisymmetric Maxwell-tensor $F_{\mu\nu}$ is replaced by the three complex scalars
\begin{equation}
\begin{aligned}{}
\phi_{0}&=F_{13}=F_{\mu\nu}l^{\mu}m^{\nu},\\
\phi_{1}&=\frac{1}{2}(F_{12}+F_{43})=\frac{1}{2}F_{\mu\nu}(l^{\mu}n^{\nu}+\xbar{m}^{\mu}m^{\nu}),\\
\phi_{2}&=F_{42}=F_{\mu\nu}\xbar{m}^{\mu}n^{\nu}.
\end{aligned}
\end{equation}
Correspondingly, the Maxwell equations in tetrad form
\begin{equation}
F_{[ab|c]}=0,\quad \eta^{nm}F_{an|m}=0,
\end{equation}
can be replaced by those equations
\begin{equation}
\begin{aligned}{}
\phi_{1|1}-\phi_{0|4}=0,&&\phi_{2|1}-\phi_{1|4}=0,\\
\phi_{1|3}-\phi_{0|2}=0,&&\phi_{2|3}-\phi_{1|2}=0.
\end{aligned}
\end{equation}
Expanding these equations in the terms of the ordinary derivatives and spinor coefficients, then expressing them in the symbols above, we can get the Maxwell equations in NP formalism. Similar disposition can be used to deal with the Klein-Gordon equation for scalar field, where we define $\Omega_{1}=D\varphi$, $\Omega_{2}=\Delta\varphi$, $\Omega=\delta\varphi$, $\bar{\Omega}=\xbar\delta\varphi$.

The freedom of the rotation of the basis vectors, see e.g. in \cite{NP}, will allow us to set
\be
\pi=\kappa=\epsilon=0,\,\,\rho=\bar\rho,\,\,\tau=\bar\alpha+\beta.
\ee
From those conditions, one can find that $l$ is tangent to a null geodesic with an affine parameter. Also, the congruence of the null geodesic is hypersurface orthogonal, that is, $l$ is proportional to the gradient of a scalar field. So it is convenient to choose the scalar field as coordinate $u=x^{1}$, and the affine parameter as $r=x^{2}$. Thus, the basis vectors and the co-tetrad must have the form
\be\begin{split}
\label{eq111}
&n^{\mu}\partial_{\mu}=\frac{\p}{\p u} + U \frac{\p}{\p r} + X^A \frac{\p}{\p x^A},\;\;\;\;\;\;l^{\mu}\partial_{\mu}=\frac{\p}{\p r},\;\;\;\;\;\;m^{\mu}\partial_{\mu}=\omega\frac{\p}{\p r} + L^A \frac{\p}{\p x^A},\\
&n_{\mu}dx^{\mu}=\big[-U-X^A(\xbar\omega L_A+\omega \bar L_A)\big]du + dr + (\omega\bar L_A+\xbar\omega L_A)dx^A,\\
&l_{\mu}dx^{\mu}=du,\;\;\;\;\;\;m_{\mu}dx^{\mu}=-X^A L_A du + L_A dx^A.
\end{split}\ee
where $L_AL^A=0$, $L_A\bar L^A=-1$. We will use the standard stereographic coordinates $z=e^{i\phi}\cot\frac {\theta}{2}$ and $\bar z=e^{-i\phi}\cot\frac{\theta}{2}$ in this work.

The Lagrangian of four-dimensional Einstein-Maxwell-dilaton theories are\footnote{Note that we use the signature $(+,-,-,-)$. Hence the third term in the Lagrangian is $\frac{1}{2}(\partial\varphi)^{2}$ rather than most used convention $-\frac{1}{2}(\partial\varphi)^{2}$, see e.g. in\cite{LMW}}
\be\label{lagrangian}
\cL=\sqrt{-g}\left[ R - \frac14 e^{a\varphi} F^2+\frac{1}{2}(\partial\varphi)^{2}\right],\qquad
F=dA.
\ee
This class of theories is generalized from the Einstein-Maxwell theory to include a real dilatonic scalar. When the dilaton coupling constant $a$ takes the following specific values $a=0,\frac{1}{\sqrt{3}},1,\sqrt{3}$, the EMD theories can be embedded into $\mathscr{N}=2$ $D=4$ STU supergravity, with global $SL(2,\mathbb R)\times SL(2,\mathbb R)\times SL(2,\mathbb R)$ triality symmetries that correspond to the string/weak, T and U dualities of string theory [].  Einstein-Maxwell theory, which is the bosonic sector of $\mathscr{N}=2$ supergravity, can be reduced from the $a=0$ case, while the $a=\sqrt{3}$ case can be Kaluza-Klein theory. Now we suppose that $a$ is an arbitrary real constant.

The dilaton, Maxwell and Einstein equations can be derived from the Lagrangian (\ref{lagrangian})
\be
\partial_{\mu}(\sqrt{-g}g^{\mu\nu}\partial_{\nu}\varphi)+\frac{a}{4}\sqrt{-g}e^{a\varphi}F^{2}=0.
\ee
\be
\partial_{\nu}(\sqrt{-g}e^{a\varphi}F^{\mu\nu})=0.
\ee
\be
R_{\mu\nu}=\frac{1}{2}e^{a\varphi}F_{\mu\rho}F_{\nu}^{\ \rho}-\frac{1}{8}g_{\mu\nu}e^{a\varphi}F^{2}-\frac{1}{2}\partial_{\mu}\varphi\partial_{\nu}\varphi.
\ee
According to these equations of motion in the tensorial form, we can easily recast them into the NP formalism. We divide these equations into three groups\cite{NU}:

I. Radial equations\\
\indent This group of equations can be integrated to find the radial dependence of all the variables, up to an proper order of magnitude. Each integration gives an arbitrary function of three nonradial coordinates (integration constant).

II. Nonradial equations\\
\indent This group of equations give the relations among these integration constants so that most of the functions can be expressed in terms of two basic functions $\sigma^{0}(u,z,\bz)$ and $P(u,z,\bz)$.

III. The u-derivative equations\\
\indent This group of equations characterizes the propagation of the components of Weyl tensor, the dilatonic scalar field and the Maxwell fields off the hypersurface in the u-direction (time direction), from null surface to null surface.

\textbf{Radial equations}
\label{NPequations}
\bea
&&D\rho =\rho^2+\sigma\xbar\sigma + \frac{1}{2}e^{a\varphi}\phi_0 \xbar\phi_0+\frac{1}{4}(\Omega_{1})^{2},\label{R1}\\
&&D\sigma=2\rho \sigma + \Psi_{0},\label{R2}\\
&&D\tau =\tau \rho +  \xbar \tau \sigma   + \Psi_1 + \frac{1}{2}e^{a\varphi}\phi_0 \xbar\phi_1+\frac{1}{4}\Omega_{1}\Omega,\label{R3}\\
&&D\alpha=\rho  \alpha + \beta \xbar \sigma  + \frac{1}{2}e^{a\varphi}\phi_1 \xbar\phi_0+\frac{1}{4}\Omega_{1}\bar{\Omega},\label{R4}\\
&&D\beta  =\alpha \sigma + \rho  \beta + \Psi_{1},\label{R5}\\
&&D\gamma=\tau \alpha +  \xbar \tau \beta  + \Psi_2 + \frac{1}{2}e^{a\varphi}\phi_1 \xbar\phi_1+\frac{1}{6}\Omega_{1}\Omega_{2}+\frac{1}{12}\Omega\bar{\Omega},\label{R6}\\
&&D\lambda=\rho\lambda + \xbar\sigma\mu + \frac{1}{2}e^{a\varphi}\phi_2 \xbar\phi_0+\frac{1}{4}(\bar{\Omega})^{2},\label{R7}\\
&&D\mu =\rho \mu + \sigma\lambda + \Psi_{2}+\frac{1}{12}(\Omega\bar{\Omega}-\Omega_{1}\Omega_{2}) ,\label{R8}\\
&&D\nu =\xbar\tau \mu + \tau  \lambda + \Psi_3 + \frac{1}{2}e^{a\varphi}\phi_2 \xbar\phi_1+\frac{1}{4}\Omega_{2}\bar{\Omega},\label{R9}\\
&&DU=\xbar\tau\omega+\tau\xbar\omega - (\gamma+\xbar\gamma),\label{R10}\\
&&DX^A=\xbar\tau L^A + \tau\bar L^A,\label{R11}\\
&&D\omega=\rho\omega+\sigma\xbar\omega-\tau,\label{R12}\\
&&DL^A=\rho L^A + \sigma \bar L^A,\label{R13}\\
&&D\Psi_1 - \xbar\delta \Psi_0 =  4 \rho \Psi_1 - 4\alpha \Psi_0+\frac{1}{2}e^{a\varphi}[(a\xbar\phi_{1}-\frac{a}{2}e^{a\varphi}\xbar\phi_{1}-\frac{a}{2}e^{a\varphi}\phi_{1})\phi_{0}\Omega_{1}\cr
&&-(a\xbar\phi_{0}-\frac{a}{2}e^{a\varphi}\xbar\phi_{0}-\frac{a}{2}e^{a\varphi}\phi_{0})\phi_{0}\Omega+\xbar\phi_{1}D\phi_{0}-\xbar\phi_{0}\delta\phi_{0}-2\sigma\phi_{1}\xbar\phi_{0}\cr
&&+2\beta\phi_{0}\xbar\phi_{0}]-\frac{1}{2}\Omega_{1}\delta\Omega_{1}+\frac{1}{4}\Omega D\Omega_{1}+\frac{1}{4}\Omega_{1}D\Omega-\frac{1}{2}\rho\Omega_{1}\Omega-\frac{1}{2}\sigma\Omega_{1}\bar{\Omega}\cr
&&+\frac{1}{2}(\xbar{\alpha}+\beta)(\Omega_{1})^{2},\label{R14}\\
&&D\Psi_2 - \xbar\delta \Psi_1 =   3\rho \Psi_2  - 2 \alpha \Psi_1- \lambda \Psi_0\cr
&&+\frac{1}{2}e^{a\varphi}[(a\xbar\phi_{1}-\frac{1}{2}a e^{a\varphi}\xbar\phi_{1}+\frac{1}{2}a e^{a\varphi}\phi_{1})\phi_{0}\bar{\Omega}-(a\xbar\phi_{0}-\frac{1}{2}a e^{a\varphi}\xbar\phi_{0})\phi_{0}\Omega_{2}\cr
&&-\frac{1}{2}a e^{a\varphi}\phi_{2}\phi_{0}\Omega_{1}+\xbar\phi_{1}\xbar\delta\phi_{0}-\xbar\phi_{0}\Delta\phi_{0}-2\alpha\phi_{0}\xbar\phi_{1}+2\rho\phi_{1}\xbar\phi_{1}+2\gamma\phi_{0}
\xbar\phi_{0}\cr
&&-2\tau\phi_{1}\xbar\phi_{0}]+\frac{1}{4}\Omega\xbar\delta\Omega_{1}+\frac{1}{4}\Omega_{1}\xbar\delta\Omega-\frac{1}{2}\Omega_{1}\Delta\Omega_{1}-\frac{1}{2}(\alpha+\xbar\tau)\Omega_{1}\Omega\cr
&&+\frac{1}{4}\rho(\Omega_{1}\Omega_{2}+\Omega\bar{\Omega})+\frac{1}{4}\xbar\sigma(\Omega)^{2}-\frac{1}{4}(\xbar\mu-2\gamma-2\xbar\gamma)(\Omega_{1})^{2}-\frac{1}{2}\tau\Omega_{1}\bar{\Omega}\cr
&&+\frac{1}{12}(\Omega_{2}D\Omega_{1}+\Omega_{1}D\Omega_{2}-\bar{\Omega}D\Omega-\Omega D\bar{\Omega}),\label{R15}\\
&&D\Psi_3 - \xbar\delta \Psi_2 =  2\rho \Psi_3 - 2\lambda \Psi_1 +\frac{1}{2}e^{a\varphi}[(a\xbar\phi_{1}-\frac{1}{2}ae^{a\varphi}\phi_{1}-\frac{1}{2}a e^{a\varphi}\xbar\phi_{1})\phi_{2}\Omega_{1}\cr
&&-(a\xbar\phi_{0}-\frac{1}{2}a e^{a\varphi}\phi_{0}-\frac{1}{2}a e^{a\varphi}\xbar\phi_{0})\phi_{2}\Omega+\xbar\phi_{1}D\phi_{2}-\xbar\phi_{0}\delta\phi_{2}+2\mu\phi_{1}\xbar\phi_{0}\cr
&&-2\beta\phi_{2}\xbar\phi_{0}]+\frac{1}{4}(\bar{\Omega}D\Omega_{2}+\Omega_{2}D\bar{\Omega})-\frac{1}{2}\bar{\Omega}\delta\bar{\Omega}-\frac{1}{2}\rho\Omega_{2}\bar{\Omega}+\frac{1}{2}\mu\Omega_{1}\bar{\Omega}\cr
&&+\frac{1}{2}(\xbar\alpha-\beta)(\bar{\Omega})^{2}-\frac{1}{12}(\Omega_{2}\xbar\delta\Omega_{1}+\Omega_{1}\xbar\delta\Omega_{2}-\bar{\Omega}\xbar\delta\Omega-\Omega\xbar\delta\bar{\Omega}),\label{R16}\\
&&D\Psi_4 - \xbar\delta \Psi_3 = \rho  \Psi_4 + 2 \alpha \Psi_3 - 3 \lambda \Psi_2+\frac{1}{2}e^{a\varphi}[(\frac{1}{2}a e^{a\varphi}\xbar\phi_{0}-a\xbar\phi_{0})\phi_{2}\Omega_{2}\cr
&&-\frac{1}{2}a e^{a\varphi}\phi_{2}\phi_{2}\Omega_{1}-(\frac{1}{2}a e^{a\varphi}\xbar\phi_{1}-\frac{1}{2}a e^{a\varphi}\phi_{1}-a\xbar\phi_{1})\phi_{2}\bar{\Omega}\cr
&&-\xbar\phi_{0}\Delta\phi_{2}+\xbar\phi_{1}\xbar\delta\phi_{2}+2\alpha\xbar\phi_{1}\phi_{2}+2\nu\phi_{1}\xbar\phi_{0}-2\gamma\xbar\phi_{0}\phi_{2}-2\lambda\phi_{1}\xbar\phi_{1}]\cr
&&+\frac{1}{2}\bar{\Omega}\Delta\bar{\Omega}-\frac{1}{4}\bar{\Omega}\xbar\delta\Omega_{2}-\frac{1}{4}\Omega_{2}\xbar\delta\bar{\Omega}-\frac{1}{2}(\alpha-\xbar\tau)\Omega_{2}\bar{\Omega}\cr
&&-\frac{1}{2}\nu\Omega_{1}\bar{\Omega}-\frac{1}{4}\xbar\sigma(\Omega_{2})^{2}+\frac{1}{4}\lambda(\Omega_{1}\Omega_{2}+\Omega\bar{\Omega})+\frac{1}{4}(\xbar\mu+2\gamma-2\xbar\gamma)(\bar{\Omega})^{2},\label{R17}\\
&&D\phi_1 - \xbar\delta \phi_0 = 2\rho \phi_1 - 2\alpha \phi_0-\frac{1}{2}a e^{a\varphi}(\phi_{1}+\xbar\phi_{1})\Omega_{1}+\frac{1}{2}a e^{a\varphi}\phi_{0}\bar{\Omega}\cr
&&\hspace{1.5cm} +\frac{1}{2} a e^{a\varphi}\xbar\phi_{0}\Omega,\label{R18}\\
&&D\phi_2 - \xbar\delta \phi_1 = \rho \phi_2 - \lambda \phi_0+\frac{1}{2}a e^{a\varphi}\xbar \phi_{0}\Omega_{2}-\frac{1}{2} a e^{a\varphi}\phi_{2}\Omega_{1}\cr
&&\hspace{1.5cm}+\frac{1}{2}a e^{a\varphi}(\phi_{1}-\xbar\phi_{1})\bar{\Omega}.\label{R19}
\eea

\textbf{Non-radial  equations}
\bea
&&\Delta\lambda  = \xbar\delta\nu- (\mu + \xbar\mu)\lambda - (3\gamma - \xbar\gamma)\lambda + 2\alpha \nu - \Psi_4,\label{H1}\\
&&\Delta\rho= \xbar\delta\tau- \rho\xbar\mu - \sigma\lambda  -2\alpha \tau + (\gamma + \xbar\gamma)\rho  - \Psi_2+\frac{1}{12}\Omega_{1}\Omega_{2}-\frac{1}{12}\Omega\bar{\Omega} ,\label{H2}\\
&&\Delta\alpha = \xbar\delta\gamma +\rho \nu - (\tau + \beta)\lambda + (\xbar\gamma - \gamma -\xbar \mu)\alpha  -\Psi_3 ,\label{H3}\\
&&\Delta \mu=\delta\nu-\mu^2 - \lambda\xbar\lambda - (\gamma + \xbar\gamma)\mu   + 2 \beta \nu - \frac{1}{2}e^{a\varphi}\phi_2 \xbar\phi_2-\frac{1}{4}(\Omega_{2})^{2},\label{H4}\\
&&\Delta \beta=\delta\gamma - \mu\tau + \sigma\nu + \beta(\gamma - \xbar\gamma -\mu) - \alpha\xbar\lambda -\frac{1}{2}e^{a\varphi} \phi_1 \xbar\phi_2-\frac{1}{4}\Omega_{2}\Omega,\label{H5}\\
&&\Delta \sigma=\delta\tau - \sigma\mu - \rho\xbar\lambda - 2 \beta \tau + (3\gamma - \xbar\gamma)\sigma  - \frac{1}{2}e^{a\varphi}\phi_0 \xbar\phi_2-\frac{1}{4}(\Omega)^{2},\label{H6}\\
&&\Delta \omega=\delta U +\xbar\nu -\xbar\lambda\xbar\omega + (\gamma-\xbar\gamma-\mu)\omega,\label{H7}\\
&&\Delta L^A=\delta X^A - \xbar\lambda \bar L^A + (\gamma-\xbar\gamma-\mu)L^A,\label{H8}\\
&&\delta\rho - \xbar\delta\sigma=\rho\tau - \sigma (3\alpha - \xbar\beta)   - \Psi_1 + \frac{1}{2}e^{a\varphi}\phi_0 \xbar\phi_1+\frac{1}{4}\Omega_{1}\Omega,\label{H9}\\
\cr
&&\delta\alpha - \xbar\delta\beta=\mu\rho - \lambda\sigma + \alpha\xbar\alpha + \beta\xbar\beta - 2 \alpha\beta - \Psi_2  + \frac{1}{2}e^{a\varphi}\phi_1 \xbar\phi_1\cr
&&+\frac{1}{12}\Omega_{1}\Omega_{2}+\frac{1}{6}\Omega\bar{\Omega},\label{H10}\\
&&\delta\lambda - \xbar\delta\mu= \mu \xbar\tau + \lambda (\xbar\alpha - 3\beta) - \Psi_3 + \frac{1}{2}e^{a\varphi}\phi_2 \xbar\phi_1+\frac{1}{4}\Omega_{2}\bar{\Omega},\label{H11}\\
&&\delta \xbar\omega-\bar\delta\omega=\mu - \xbar\mu - (\alpha - \xbar\beta) \omega +  (\xbar\alpha - \beta)\xbar\omega,\label{H12}\\
&&\delta \bar L^A - \bar\delta L^A= (\xbar\alpha - \beta)\bar L^A -  (\alpha - \xbar\beta) L^A.\label{H13}
\eea

\textbf{The u-derivative equations}
\bea
&&\Delta\Psi_0 - \delta \Psi_1 = (4\gamma -\mu)\Psi_0 - (4\tau + 2\beta)\Psi_1 + 3\sigma \Psi_2\cr
&&-\frac{1}{2}e^{a\varphi}[(a\xbar\phi_{2}-\frac{1}{2}a e^{a\varphi}\xbar\phi_{2})\phi_{0}\Omega_{1}+\frac{1}{2}a e^{a\varphi}\phi_{0}\phi_{0}\Omega_{2}+\cr
&&(\frac{1}{2}a e^{a\varphi}\xbar\phi_{1}-\frac{1}{2}a e^{a\varphi}\phi_{1}-a\xbar\phi_{1})\phi_{0}\Omega+\xbar\phi_{2}D\phi_{0}-\xbar\phi_{1}\delta\phi_{0}+2\beta\phi_{0}\xbar\phi_{1}\cr
&&-2\sigma\phi_{1}\xbar\phi_{1}]-\frac{1}{2}\Omega D\Omega+\frac{1}{4}\Omega\delta\Omega_{1}+\frac{1}{4}\Omega_{1}\delta\Omega-\frac{1}{2}\beta\Omega_{1}\Omega-\frac{1}{4}\xbar\lambda(\Omega_{1})^{2}\cr
&&+\frac{1}{4}\sigma(\Omega_{1}\Omega_{2}+\Omega\bar{\Omega})+\frac{1}{4}\rho(\Omega)^{2}.\label{H14}\\
&&\Delta\Psi_1 - \delta \Psi_2 = \nu\Psi_0 + (2\gamma - 2\mu)\Psi_1 - 3\tau \Psi_2 + 2\sigma \Psi_3 \nn\\
&&+\frac{1}{2}e^{a\varphi}[(a\xbar\phi_{1}-\frac{1}{2}a e^{a\varphi}\xbar\phi_{1}-\frac{1}{2}a e^{a\varphi}\phi_{1})\phi_{0}\Omega_{2}+(\frac{1}{2}a e^{a\varphi}\phi_{2}+\frac{1}{2}a e^{a\varphi}\xbar\phi_{2})\phi_{0}\Omega\cr
&&-a\phi_{0}\xbar\phi_{2}\bar{\Omega}+\xbar\phi_{1}\Delta\phi_{0}-\xbar\phi_{2}\xbar\delta\phi_{0}-2\gamma\phi_{0}\xbar\phi_{1}-2\rho\phi_{1}\xbar\phi_{2}+2\alpha\phi_{0}\xbar\phi_{2}+2\tau\phi_{1}\xbar\phi_{1}]\cr
&&+\frac{1}{4}\Omega\Delta\Omega_{1}+\frac{1}{4}\Omega_{1}\Delta\Omega-\frac{1}{2}\Omega\xbar\delta\Omega+\frac{1}{2}(\xbar\mu-\gamma)\Omega_{1}\Omega-\frac{1}{2}\rho\Omega_{2}\Omega-\frac{1}{4}\xbar\nu(\Omega_{1})^{2}\cr
&&+\frac{1}{4}\tau(\Omega_{1}\Omega_{2}+\Omega\bar{\Omega})+\frac{1}{4}(\xbar\tau-2\xbar\beta+2\alpha)(\Omega)^{2}-\frac{1}{12}(\Omega_{2}\delta\Omega_{1}+\Omega_{1}\delta\Omega_{2}\cr
&&-\Omega\delta\bar{\Omega}-\bar{\Omega}\delta\Omega).\label{H15}\\
&&\Delta\Psi_2 - \delta \Psi_3 = 2\nu \Psi_1 - 3\mu \Psi_2 + (2\beta - 2\tau) \Psi_3 + \sigma \Psi_4\cr
&&-\frac{1}{2}e^{a\varphi}[(a\xbar\phi_{2}-\frac{1}{2}a e^{a\varphi}\xbar\phi_{2})\phi_{2}\Omega_{1}+\frac{1}{2}a e^{a\varphi}\phi_{0}\phi_{2}\Omega_{2}\cr
&&+(\frac{1}{2}a e^{a\varphi}\xbar\phi_{1}-\frac{1}{2}a e^{a\varphi}\phi_{1}-a\xbar\phi_{1})\phi_{2}\Omega+\xbar\phi_{2}D\phi_{2}-\xbar\phi_{1}\delta\phi_{2}-2\beta\xbar\phi_{1}\phi_{2}\cr
&&+2\mu\xbar\phi_{1}\phi_{1}]-\frac{1}{2}\Omega_{2}D\Omega_{2}+\frac{1}{4}\bar{\Omega}\delta\Omega_{2}+\frac{1}{4}\Omega_{2}\delta\bar{\Omega}+\frac{1}{2}\beta\Omega_{2}\bar{\Omega}\cr
&&-\frac{1}{4}\mu(\Omega_{1}\Omega_{2}+\Omega\bar{\Omega})-\frac{1}{4}\xbar\lambda(\bar{\Omega})^{2}+\frac{1}{4}\rho(\Omega_{2})^{2}+\frac{1}{12}(\Omega_{2}\Delta\Omega_{1}+\Omega_{1}\Delta\Omega_{2}\cr
&&-\bar{\Omega}\Delta\Omega-\Omega\Delta\bar{\Omega}).\label{H16}\\
&&\Delta\Psi_3 - \delta \Psi_4 = 3\nu \Psi_2 - (2\gamma + 4\mu) \Psi_3 + (4\beta - \tau) \Psi_4 \nn\\
&&+\frac{1}{2}e^{a\varphi}[(a\xbar\phi_{1}-\frac{1}{2}a e^{a\varphi}\xbar\phi_{1}-\frac{1}{2}a e^{a\varphi}\phi_{1})\phi_{2}\Omega_{2}-(a\phi_{2}-\frac{1}{2}a e^{a\varphi}\phi_{2})\xbar\phi_{2}\bar{\Omega}\cr
&&+\frac{1}{2}a\phi_{2}\phi_{2}e^{a\varphi}\Omega+\xbar\phi_{1}\Delta\phi_{2}-\xbar\phi_{2}\xbar\delta\phi_{2}+2\gamma\xbar\phi_{1}\phi_{2}-2\nu\phi_{1}\xbar\phi_{1}+2\lambda\phi_{1}\xbar\phi_{2}\cr
&&-2\alpha\phi_{2}\xbar\phi_{2}]+\frac{1}{4}\bar{\Omega}\Delta\Omega_{2}+\frac{1}{4}\Omega_{2}\Delta\bar{\Omega}-\frac{1}{2}\Omega_{2}\xbar\delta\Omega_{2}+\frac{1}{2}(\xbar\mu+\gamma)\Omega_{2}\bar{\Omega}\cr
&&-\frac{1}{4}\nu(\Omega_{1}\Omega_{2}+\Omega\bar{\Omega})-\frac{1}{4}\xbar\nu(\bar{\Omega})^{2}+\frac{1}{2}\lambda\Omega_{2}\Omega-\frac{1}{4}(\alpha+\xbar\beta)(\Omega_{2})^{2}.\label{H17}\\
&&\Delta\phi_0 - \delta \phi_1 = (2\gamma-\mu) \phi_0 - 2\tau \phi_1 + \sigma \phi_2-\frac{1}{2}a e^{a\varphi}\phi_{0}\Omega_{2}+\frac{1}{2}a e^{a\varphi}\xbar\phi_{2}\Omega_{1}\cr
&&+\frac{1}{2}a e^{a\varphi}\phi_{1}\Omega-\frac{1}{2}a e^{a\varphi}\xbar\phi_{1}\Omega.\label{H18}\\
&&\Delta\phi_1 - \delta \phi_2 = \nu \phi_0 - 2 \mu \phi_1 - (\xbar\alpha - \beta) \phi_2-\frac{1}{2}a e^{a\varphi}\phi_{1}\Omega_{2}-\frac{1}{2}a e^{a\varphi}\xbar\phi_{1}\Omega_{2}\cr
&&+\frac{1}{2}a e^{a\varphi}\xbar\phi_{2}\bar{\Omega}+\frac{1}{2}a e^{a\varphi}\phi_{2}\Omega.\label{H19}\\
&&\Delta\Omega_{1}+D\Omega_{2}=\xbar\delta\Omega+\delta\bar{\Omega}+(\gamma+\xbar\gamma-\mu-\xbar\mu)\Omega_{1}+2\rho\Omega_{2}-2\xbar\alpha\bar{\Omega}\cr
&&\hspace{1.5cm}-2\alpha\Omega+a e^{a\varphi}(\phi_{1}^{2}+\xbar\phi_{1}^{2}-\xbar\phi_{0}\xbar\phi_{2}-\phi_{0}\phi_{2}).\label{H20}
\eea

\section{The solution space}
The main condition of approaching flatness at infinity is $\Psi^{0}=\frac{\Psi_{0}^{0}}{r^{5}}+\cO(r^{-6})$. Newman and Unti\cite{NU} listed the fall-off conditions of the rest quantities by solving the empty space Newman-Penrose equations
\bea
&&\rho=-r^{-1}+\cO(r^{-3}),\;\;\sigma=\cO(r^{-2}),\;\;\alpha=\cO(r^{-1}),\;\;\beta=\cO(r^{-1}),\cr
&&\tau=\cO(r^{-3}),\;\;\lambda=\cO(r^{-1}),\;\;\mu=\cO(r^{-1}),\;\;\gamma=\cO(1),\cr
&&\nu=\cO(1),\;\;U=\cO(r),\;\;X^{z}=\cO(r^{-3}),\;\;\omega=\cO(r^{-1}),\cr
&&L^{z}=\cO(r^{-2}),\;\;L^{\bz}=\cO(r^{-1}),\cr
&&\Psi_{1}=\cO(r^{-4}),\;\;\Psi_{2}=\cO(r^{-3}),\;\;\Psi_{3}=\cO(r^{-2}),\;\;\Psi_{4}=\cO(r^{-1}).
\eea
The fall-off of the matter fields can not violate the asymptotic conditions above, so we choose
\begin{equation}
\label{eqs2}
\phi_{0}=\frac{\phi_{0}^{0}}{r^{3}}+\cO(r^{-4}),
\end{equation}
\begin{equation}
\label{eqs3}
\varphi=\frac{\varphi_{1}}{r}+\frac{\varphi_{2}}{r^{2}}+\cO(r^{-3}),
\end{equation}
and
\begin{equation}
\phi_{1}=\frac{\phi_{1}^{0}}{r^{2}}+\cO(r^{-3}),\quad \phi_{2}^{0}=\frac{\phi_{2}^{0}}{r}+\cO(r^{-2}).
\end{equation}
Using the conditions above, we work out the asymptotically flat solution space of Newman-Penrose equations in Einstein-Maxwell-dilaton theories. The solutions of the radial equations are as follows. One should notice that here the boundary topology is an arbitrary 2 surface but not $S^{2}$. Here the ``$\eth$'' operator is defined as
\begin{equation}\begin{split}
&\eth \eta^s=P\bP^{-s}\p_{\bz}(\bP^s \eta^s)=P\p_{\bz} \eta^s + 2 s\xbar\alpha^0 \eta^s,\\
&\xbar\eth \eta^s=\bP P^{s}\p_z(P^{-s} \eta^s)=\bP\p_z \eta^s -2 s \alpha^0 \eta^s.
\end{split}\end{equation}
where $s$ is the spin weight of the field $\eta$. The spin weights of relevant fields are listed in Table \ref{t1}.
\begin{table}[h]
\caption{Spin weights}\label{t1}
\begin{center}\begin{tabular}{|c|c|c|c|c|c|c|c|c|c|c|c|c|c|c|c|c|c}
\hline
& $\eth$ & $\p_u$ & $\gamma^0$ & $\nu^0$ & $\mu^0$ & $\sigma^0$ & $\lambda^0$  & $\Psi^0_4$ &  $\Psi^0_3$ & $\Psi^0_2$ & $\Psi^0_1$ & $\Psi_0^0$ &$\phi^0_2$ & $\phi^0_1$ & $\phi_0^0$   \\
\hline
s & $1$& $0$& $0$& $-1$& $0$& $2$& $-2$  &
  $-2 $&  $-1$ & $0$ & $1$ & $2$  & $-1$ & $0$ & $1$   \\
\hline
\end{tabular}\end{center}\end{table}
\begin{equation}
\begin{aligned}{}
\Psi_0&=\frac{\Psi_0^0(u,z,\bz)}{r^5} + \frac{\Psi_0^1(u,z,\bz)}{r^6} + \cO(r^{-7}),\\
\phi_0&=\frac{\phi_0^0(u,z,\bz)}{r^3} + \frac{\phi_0^1(u,z,\bz)}{r^4} + \cO(r^{-5}),\\
\varphi&=\frac{\varphi_{1}}{r}+\frac{\varphi_{2}}{r^{2}}+\cO(r^{-3}),\\
\rho&=-\frac{1}{r}+\frac{-\varphi_{1}^{2}-4\sigma^{0}\xbar\sigma^{0}}{4 r^{3}}-\frac{\varphi_{1}\varphi_{2}}{2r^{4}}+\frac{1}{48r^{5}}(-8\phi_{0}^{0}\xbar\phi_{0}^{0}-\varphi_{1}^{4}-16\varphi_{2}^{2}-24\varphi_{1}\varphi_{3}\\
&+8\xbar\sigma^{0}\Psi_{0}^{0}+8\sigma^{0}\xbar\Psi_{0}^{0}-16\sigma^{0}\xbar\sigma^{0}\varphi_{1}^{2}-48(\sigma\xbar\sigma)^{2})+\cO(r^{-6}),\\
\sigma&=\frac{\sigma^{0}(u,z,\xbar z)}{r^{2}}+\frac{-2\Psi^{0}_{0}+\sigma^{0}\varphi_{1}^{2}+4\sigma^{0}\sigma^{0}\xbar\sigma^{0}}{4 r^{4}}+\frac{1}{3r^{5}}(-\Psi^{1}_{0}+\sigma^{0}\varphi_{1}\varphi_{2})+\cO(r^{-6}),\\
L^{z}&=-\frac{\sigma^{0}\bP(u,z,\bz)}{r^{2}}+\frac{1}{24 r^{4}}(4\bP\Psi^{0}_{0}-5\bP\sigma^{0}\varphi_{1}^{2}-24\bP \xbar\sigma^{0}(\sigma^{0})^{2})+\frac{1}{12 r^{5}}(\bP\Psi_{0}^{1}\\
&-3\bP\varphi_{1}\varphi_{2}\sigma^{0})+\cO(r^{-6}),\\
L^{\bz}&=\frac{P(u,z,\bz)}{r}+\frac{1}{8r^{3}}(P\varphi_{1}^{2}+8P\sigma^{0}\xbar\sigma^{0})+\frac{1}{6r^{4}}P\varphi_{1}\varphi_{2}+\frac{1}{384 r^{5}}(16 P\phi_{0}^{0}\xbar\phi_{0}^{0}\\
&+5P\varphi_{1}^{4}+32P\varphi_{2}^{2}+48P\varphi_{1}\varphi_{3}-64P\xbar\sigma^{0}\Psi_{0}^{0}-32P\sigma^{0}\xbar\Psi_{0}^{0}+112P\xbar\sigma^{0}\sigma^{0}\varphi_{1}^{2}\cr
&+384P(\xbar\sigma^{0})^{2}(\sigma^{0})^{2})+\cO(r^{-6}),\\
L_{z}&=-\frac{r}{\bP}+\frac{\varphi_{1}^{2}}{8\bP r}+\frac{\varphi_{1}\varphi_{2}}{6\bP r^{2}}+\frac{1}{384\bP r^{3}}(16\phi_{0}^{0}\xbar\phi_{0}^{0}-\varphi_{1}^{4}+32\varphi_{2}^{2}+48\varphi_{1}\varphi_{3}+32\xbar\sigma^{0}\Psi_{0}^{0})\\
&+\cO(r^{-4}),\\
L_{\bz}&=-\frac{\sigma^{0}}{P}+\frac{1}{24Pr^{2}}(4\Psi_{0}^{0}+\sigma^{0}\varphi_{1}^{2})+\frac{1}{12 P r^{3}}(\Psi_{0}^{1}+\sigma^{0}\varphi_{1}\varphi_{2})+\cO(r^{-4}),\\
\alpha&=\frac{\alpha^{0}}{r}+\frac{\xbar\sigma^{0}\xbar\alpha^{0}}{r^{2}}+\frac{1}{8r^{3}}(\alpha^{0}\varphi_{1}^{2}+8\alpha^{0}\xbar\sigma^{0}\sigma^{0}+\varphi_{1}\xbar\eth\varphi_{1})\\
&+\frac{1}{24 r^{4}}(24\xbar\alpha^{0}\sigma^{0}(\xbar\sigma^{0})^{2}-4\xbar\alpha^{0}\xbar\Psi^{0}_{0}+4\xbar\sigma^{0}\Psi_{1}^{0}-4\phi_{1}^{0}\xbar\phi_{0}^{0}-2\xbar\omega^{0}\varphi_{1}^{2}+4\alpha^{0}\varphi_{1}\varphi_{2}\\
&+5\xbar\alpha^{0}\xbar\sigma^{0}(\varphi_{1})^{2}-2\xbar\sigma^{0}\varphi_{1}\eth\varphi_{1}+4\varphi_{2}\xbar\eth\varphi_{1}+2\varphi_{1}\xbar\eth\varphi_{2})+\cO(r^{-5}),\\
\beta&=-\frac{\xbar\alpha^{0}}{r}-\frac{\alpha^{0}\sigma^{0}}{r^{2}}+\frac{1}{8r^{3}}(-\xbar\alpha^{0}\varphi_{1}^{2}-4\Psi_{1}^{0}-8\sigma^{0}\xbar\sigma^{0}\xbar\alpha^{0})\\
&+\frac{1}{24 r^{4}}(8\xbar\eth\Psi_{0}^{0}+4\alpha^{0}\Psi_{0}^{0}-24\alpha^{0}(\sigma^{0})^{2}\xbar\sigma^{0}-12\phi_{0}^{0}\xbar\phi_{1}^{0}+2\omega^{0}\varphi_{1}^{2}-4\xbar\alpha^{0}\varphi_{1}\varphi_{2}\\
&-5\alpha^{0}\sigma^{0}\varphi_{1}^{2}+4\varphi_{2}\eth\varphi_{1}-2\varphi_{1}\eth\varphi_{2}+\sigma^{0}\varphi_{1}\xbar\eth\varphi_{1})+\cO(r^{-5}),\nonumber
\end{aligned}
\end{equation}
\begin{equation}
\begin{aligned}
\tau&=\frac{1}{8r^{3}}(-4\Psi_{1}^{0}+\varphi_{1}\eth\varphi_{1})+\frac{1}{24 r^{4}}(8\xbar\eth\Psi_{0}^{0}+4\sigma^{0}\xbar\Psi_{1}^{0}-16\phi_{0}^{0}\xbar\phi_{1}^{0}-\sigma^{0}\varphi_{1}\xbar\eth\varphi_{1}+8\varphi_{2}\eth\varphi_{1})+\cO(r^{-5}),\\
\omega&=\frac{\omega^{0}}{r}+\frac{1}{8r^{2}}(-4\Psi_{1}^{0}-8\xbar\omega^{0}\sigma^{0}+\varphi_{1}\eth\varphi_{1})+\frac{1}{24 r^{3}}(4\xbar\eth\Psi_{0}^{0}+24\sigma^{0}\xbar\sigma^{0}\omega^{0}+8\sigma^{0}\xbar\Psi^{0}_{1}-8\phi_{0}^{0}\xbar\phi_{1}^{0}\\
&+3\omega^{0}\varphi_{1}^{2}-2\sigma^{0}\varphi_{1}\xbar\eth\varphi_{1}+4\varphi_{2}\eth\varphi_{1})+\cO(r^{-4}),\\
\Psi_{1}&=\frac{\Psi_{1}^{0}(u,z,\bz)}{r^{4}}+\frac{1}{4r^{5}}(6\phi_{0}^{0}\xbar\phi_{1}^{0}-4\xbar\eth\Psi_{0}^{0}-\omega^{0}\varphi_{1}^{2}-2\varphi_{2}\eth\varphi_{1}+ \varphi_{1}\eth \varphi_{2}-\sigma^{0}\varphi_{1}\xbar\eth\varphi_{1})+\cO(r^{-6}),\\
X^{z}&=\frac{1}{24r^{3}}(4\bP\Psi_{1}^{0}-\bP\varphi_{1}\eth\varphi_{1})+\frac{1}{24r^{4}}(-2\bP\xbar\eth\Psi_{0}^{0}-4\bP\sigma^{0}\xbar\Psi_{1}^{0}+4\bP\phi_{0}^{0}\xbar\phi_{1}^{0}\\
&+\bP\sigma^{0}\varphi_{1}\xbar\eth\varphi_{1}-2\bP\varphi_{2}\eth\varphi_{1})+\cO(r^{-5}),\\
\gamma&=\gamma^{0}+\frac{1}{12r^{2}}(\xbar\gamma^{0}\varphi_{1}^{2}+\gamma^{0}\varphi_{1}^{2}-6\Psi_{2}^{0}+\varphi_{1}\partial_{u}\varphi_{1})+\frac{1}{24 r^{3}}(8\xbar\eth\Psi_{1}^{0}+4\alpha^{0}\Psi_{1}^{0}\\
&-4\xbar\alpha^{0}\xbar\Psi_{1}^{0}-12\phi_{1}^{0}\xbar\phi_{1}^{0}-2\mu^{0}\varphi_{1}^{2}+2U^{0}\varphi_{1}^{2}+4\varphi_{2}\partial_{u}\varphi_{1}-2\varphi_{1}\partial_{u}\varphi_{2}-2\xbar\eth\varphi_{1}\eth\varphi_{1}\\
&-2\varphi_{1}\xbar\eth\eth\varphi_{1}-\alpha^{0}\varphi_{1}\eth\varphi_{1}+\xbar\alpha^{0}\varphi_{1}\xbar\eth\varphi_{1})+\cO(r^{-4}),\\
\lambda&=\frac{\lambda^{0}}{r}-\frac{\xbar\sigma^{0}\mu^{0}}{r^{2}}+\frac{1}{24r^{3}}(24\sigma^{0}\xbar\sigma^{0}\lambda^{0}+12\xbar\sigma^{0}\Psi_{2}^{0}-6\phi_{2}^{0}\xbar\phi_{0}^{0}+3\lambda^{0}\varphi_{1}^{2}+(\gamma^{0}+\xbar\gamma^{0})\xbar\sigma^{0}\varphi_{1}^{2}\\
&-3(\xbar\eth\varphi_{1})^{2}+\xbar\sigma^{0}\varphi_{1}\partial_{u}\varphi_{1})+\cO(r^{-5}),\\
\mu&=\frac{\mu^{0}}{r}+\frac{1}{12 r^{2}}(-12\Psi_{2}^{0}-12\sigma^{0}\lambda^{0}+U^{0}\varphi_{1}^{2}-\varphi_{1}\partial_{u}\varphi_{1})+\frac{1}{8r^{3}}(8\sigma^{0}\xbar\sigma^{0}\mu^{0}+4\xbar\eth\Psi_{1}^{0}-4\phi_{1}^{0}\xbar\phi_{1}^{0}\\
&-4(\xbar\gamma^{0}+\gamma^{0})\varphi_{1}\varphi_{2}+2U^{0}\varphi_{1}^{2}-\xbar\eth\varphi_{1}\eth\varphi_{1}-\varphi_{1}\xbar\eth\eth\varphi_{1}-2\varphi_{1}\partial_{u}\varphi_{2})+\cO(r^{-4}),\\
U&=-r(\gamma^{0}+\xbar\gamma^{0})+U^{0}+\frac{1}{6r}(\xbar\gamma^{0}\varphi_{1}^{2}+\gamma^{0}\varphi_{1}^{2}-3\xbar\Psi_{2}^{0}-3\Psi_{2}^{0}+\varphi_{1}\partial_{u}\varphi_{1})\\
&+\frac{1}{24r^{2}}(-12\phi_{1}^{0}\xbar\phi_{1}^{0}+4\xbar\eth\Psi_{1}^{0}+4\eth\xbar\Psi_{1}^{0}-2\mu^{0}\varphi_{1}^{2} +2\varphi_{1}^{2}U^{0}+4\varphi_{2}\partial_{u}\varphi_{1}-2\varphi_{1}\partial_{u}\varphi_{2}\\
&-2\xbar\eth\varphi_{1}\eth\varphi_{1}-\varphi_{1}\xbar\eth\eth\varphi_{1}-\varphi_{1}\eth\xbar\eth\varphi_{1})+\cO(r^{-3}),\\
\Psi_{2}&=\frac{\Psi_{2}^{0}(u,z,\bz)}{r^{3}}+\frac{1}{12 r^{4}}(12\phi_{1}^{0}\xbar\phi_{1}^{0}+3\mu^{0}\varphi_{1}^{2}+8(\gamma^{0}+\xbar\gamma^{0})\varphi_{1}\varphi_{2}-12\xbar\eth\Psi_{1}^{0}-5\varphi_{1}^{2}U^{0}\\
&-2\varphi_{2}\partial_{u}\varphi_{1}+5\varphi_{1}\partial_{u}\varphi_{2}+2\xbar\eth\varphi_{1}\eth\varphi_{1}+3\varphi_{1}\xbar\eth\eth\varphi_{1})+\cO(r^{-5}),\\
\nu&=\nu^{0}-\frac{\Psi_{3}^{0}}{r}+\frac{1}{24 r^{2}}(-12\phi_{2}^{0}\xbar\phi_{1}^{0}+12\xbar\eth\Psi_{2}^{0}+\xbar\eth\xbar\gamma^{0}\varphi_{1}^{2}+\xbar\eth\gamma^{0}\varphi_{1}^{2}-4\xbar\gamma^{0}\varphi_{1}\xbar\eth\varphi_{1}\\
&-4\gamma^{0}\varphi_{1}\xbar\eth\varphi_{1}-5\xbar\eth\varphi_{1}\partial_{u}\varphi_{1}+\varphi_{1}\xbar\eth\partial_{u}\varphi_{1})+\cO(r^{-3}),\\
\Psi_{3}&=\frac{\Psi_{3}^{0}}{r^{2}}+\frac{1}{12 r^{3}}(6\phi_{2}^{0}\xbar\phi_{1}^{0}-12\xbar\eth\Psi_{2}^{0}-(\xbar\eth\gamma^{0}+\xbar\eth\xbar\gamma^{0})\varphi_{1}^{2}+(\gamma^{0}+\xbar\gamma^{0})\varphi_{1}\xbar\eth\varphi_{1}\\
&+2\xbar\eth\varphi_{1}\partial_{u}\varphi_{1}-
\varphi_{1}\xbar\eth\partial_{u}\varphi_{1})+\cO(r^{-4}),\\
\phi_{2}&=\frac{\phi_{2}^{0}(u,z,\bz)}{r}+\frac{1}{2r^{2}}(-a\phi_{2}^{0}\varphi_{1}-2\xbar\eth\phi_{1}^{0})+\frac{1}{8r^{3}}(4\lambda^{0}\phi_{0}^{0}+8\xbar\omega^{0}\phi_{1}^{0}+4\sigma^{0}\xbar\sigma^{0}\phi_{2}^{0}\\
&+4\xbar\sigma^{0}\eth\phi_{1}^{0}+4\xbar\eth^{2}\phi_{0}^{0}-2a(\gamma^{0}+\xbar\gamma^{0})\xbar\phi_{0}^{0}\varphi_{1}+\phi_{2}^{0}\varphi_{1}^{2}-a^{2}\phi_{2}^{0}\varphi_{1}^{2}-4a\phi_{2}^{0}\varphi_{2}\\
&+2a\xbar\eth\xbar\phi_{1}^{0}\varphi_{1}+4a\xbar\eth\phi_{1}^{0}\varphi_{1}+4a\xbar\phi_{1}^{0}\xbar\eth\varphi_{1}-2a\xbar\phi_{0}^{0}\partial_{u}\varphi_{1})+\cO(r^{-4}),\nonumber
\end{aligned}
\end{equation}
\begin{equation}
\begin{aligned}
\phi_{1}&=\frac{\phi_{1}^{0}(u,z,\bz)}{r^{2}}+\frac{1}{2r^{3}}(-2\xbar\eth\phi_{0}^{0}-a\phi_{1}^{0}\varphi_{1}-a\xbar\phi_{1}^{0}\varphi_{1})+\cO(r^{-4}),\\
\Psi_{4}&=\frac{\Psi_{4}^{0}}{r}-\frac{\xbar\eth\Psi_{3}^{0}}{r^{2}}+\cO(r^{-3}).
\end{aligned}
\end{equation}
The solutions of non-radial equations are as follows
\begin{equation}
\begin{aligned}
\Psi_{4}^{0}&=\xbar\eth\nu^{0}-\partial_{u}\lambda^{0}-4\gamma^{0}\lambda^{0},\\
U^{0}&=\mu^{0},\quad \gamma^{0}=-\frac{1}{2}\partial_{u}\ln\bP,\quad \omega^{0}=\xbar\eth\sigma^{0},\\
\nu^{0}&=\xbar\eth(\gamma^{0}+\xbar\gamma^{0}),\quad \alpha^{0}=\frac{1}{2}\bP\partial_{z}\ln P,\\
\lambda^{0}&=\partial_{u}\xbar\sigma^{0}+\xbar\sigma^{0}(3\gamma^{0}-\xbar\gamma^{0}),\\
\Psi_{3}^{0}&=\xbar\eth\mu^{0}-\eth\lambda^{0},\\
\mu^{0}&=-\frac{1}{2}P\bP\partial_{z}\partial_{\bz}\ln P\bP,\\
\Psi_{2}^{0}-\xbar\Psi_{2}^{0}&=\xbar\eth^{2}\sigma^{0}-\eth^{2}\xbar\sigma^{0}+\xbar\sigma^{0}\xbar\lambda^{0}-\sigma^{0}\lambda^{0}.
\end{aligned}
\end{equation}
And we obtain the solutions of the u-derivative equations which determine the propagation of the fields off the null hypersurface
\begin{equation}
\begin{aligned}{}
\partial_{u}\Psi_{0}^{0}&=\frac{3}{2}\phi_{0}^{0}\xbar\phi_{2}^{0}-\frac{1}{4}\xbar\lambda^{0}\varphi_{1}^{2}-(\gamma^{0}+5\xbar\gamma^{0})\Psi_{0}^{0}+\eth\Psi_{1}^{0}-\frac{1}{4}(\gamma^{0}+\xbar\gamma^{0})\sigma^{0}\varphi_{1}^{2}
+3\sigma^{0}\Psi_{2}^{0}\\
&-\frac{1}{4}\sigma^{0}\varphi_{1}\partial_{u}\varphi_{1}-\frac{1}{4}\varphi_{1}\eth^{2}\varphi_{1}+\frac{1}{2}(\eth\varphi_{1})^{2}.
\end{aligned}
\end{equation}
\begin{equation}
\begin{aligned}
\partial_{u}\Psi_{1}^{0}&=\phi_{1}^{0}\xbar\phi_{2}^{0}-\frac{1}{4}\xbar\nu^{0}\varphi_{1}^{2}-2(\gamma^{0}+2\xbar\gamma^{0})\Psi_{1}^{0}+2\sigma^{0}\Psi_{3}^{0}+\frac{1}{12}(\eth\xbar\gamma^{0}+\eth\gamma^{0})\varphi_{1}^{2}\\
&-\frac{1}{3}\xbar\gamma^{0}\varphi_{1}\eth\varphi_{1}+\frac{1}{6}\gamma^{0}\varphi_{1}\eth\varphi_{1}+\eth\Psi_{2}^{0}-\frac{1}{4}\varphi_{1}\partial_{u}\eth\varphi_{1}+\frac{1}{3}\eth\varphi_{1}\partial_{u}\varphi_{1}+\frac{1}{12}\varphi_{1}\eth\partial_{u}\varphi_{1}.
\end{aligned}
\end{equation}
\begin{equation}
\begin{aligned}{}
\label{mass}
\partial_{u}\Psi_{2}^{0}&=\frac{1}{2}\phi_{2}^{0}\xbar\phi_{2}^{0}-3(\gamma^{0}+\xbar\gamma^{0})\Psi_{2}^{0}+\sigma^{0}\Psi_{4}^{0}+\eth\Psi_{3}^{0}-\frac{1}{12}(\partial_{u}\gamma^{0}+\partial_{u}\xbar\gamma^{0})\varphi_{1}^{2}\\
&+\frac{1}{12}(\xbar\gamma^{0}+\gamma^{0})\varphi_{1}\partial_{u}\varphi_{1}+\frac{1}{6}(\partial_{u}\varphi_{1})^{2}-\frac{1}{12}\varphi_{1}\partial_{u}^{2}\varphi_{1}.
\end{aligned}
\end{equation}
\begin{equation}
\partial_{u}\Psi_{3}^{0}=\eth\Psi_{4}^{0}-2(2\gamma^{0}+\xbar\gamma^{0})\Psi_{3}^{0}.
\end{equation}
\begin{equation}
\label{ph00}
\partial_{u}\phi_{0}^{0}=-(\gamma^{0}+3\xbar\gamma^{0})\phi_{0}^{0}+\sigma^{0}\phi_{2}^{0}+\eth\phi_{1}^{0}-\frac{1}{2}a\xbar\phi_{2}^{0}\varphi_{1}.
\end{equation}
\begin{equation}
\label{charge}
\partial_{u}\phi_{1}^{0}=\eth\phi_{2}^{0}-2(\gamma^{0}+\xbar\gamma^{0})\phi_{1}^{0}.
\end{equation}
\begin{equation}
\partial_{u}\varphi_{2}=-2(\gamma^{0}+\xbar\gamma^{0})\varphi_{2}-\eth\xbar\eth\varphi_{1}.
\end{equation}

From the solutions above, we find that there is no constraint at the order $\cO(\frac{1}{r})$ of $\phi_{2}$ and $\varphi$, and at the order $\cO(\frac{1}{r^{2}})$ of $\sigma$. So $\sigma^{0}$, $\phi_{2}^{0}$ and $\varphi_{1}$ are related to the news functions in the system which indicate gravitational, electromagnetic and scalar radiations. $\sigma^{0}$ has a special geometric meaning that it represents the asymptotic shear of $l$ (see \cite{NP} and \cite{MW}), the change of which at early time $u_{i}$ and late time $u_{f}$ is equivalent to the time integration of the asymptotic shear of $n$, i.e. $\lambda^{0}$ when we set the boundary topology to be $S^{2}\times R$, i.e. $P=\bP=P_{s}=\frac{1+z\bz}{\sqrt{2}}$ (see \cite{F} and \cite{MW}). The memory effects\cite{F,MW}, which will be discussed in the following section, are controlled by the time integration of the asymptotic shear of $n$, i.e.$\lambda^{0}$, thus the change of the asymptotic shear at early time and late time $\Delta\sigma^{0}$ is a very important quantity that characterizes gravitational memory effects.

According to eq(\ref{ph00}), we can find that this time evolution equation involves the coupling constant $a$ which represents the non-minimal coupling of the electromagnetic field and the scalar field. $\phi_{0}^{0}$ is related to the electric dipole\cite{JN}. Our result is consistent with the result in \cite{LMW}. The phenomenon that the coupling constant $a$ do not appear in the time evolution functions of four tetrad components of Weyl tensor reflects that the scalar field is minimally coupled to gravity.

From eq(\ref{mass}) and eq(\ref{charge}) we can consider the conservation laws and the loss of mass in EMD theory. Here we work in the unit 2-sphere case. From eq(\ref{mass}), we can find
\begin{equation}
\partial_{u}\Psi_{2}^{0}=\frac{1}{2}\phi_{2}^{0}\xbar\phi_{2}^{0}-\sigma^{0}\partial_{u}^{2}\xbar\sigma^{0}-\eth^{2}\partial_{u}\xbar\sigma^{0}+\eth\Psi_{3}^{0}+\frac{1}{6}(\partial_{u}\varphi_{1})^{2}-\frac{1}{12}\varphi_{1}\partial_{u}^{2}\varphi_{1}.
\end{equation}
Define the mass density
\begin{equation}
M=\frac{1}{2}(\Psi_{2}^{0}+\xbar\Psi_{2}^{0})+\frac{1}{2}(\sigma^{0}\partial_{u}\xbar\sigma^{0}+\xbar\sigma^{0}\partial_{u}\sigma^{0})+\frac{1}{2}(\eth^{2}\xbar\sigma^{0}+\xbar\eth^{2}\sigma^{0})+\frac{1}{12}\varphi_{1}\partial_{u}\varphi_{1}.
\end{equation}
We can obtain
\begin{equation}
\partial_{u}M=\frac{1}{2}\phi_{2}^{0}\xbar\phi_{2}^{0}+\partial_{u}\sigma^{0}\partial_{u}\xbar\sigma^{0}+\frac{1}{4}(\partial_{u}\varphi_{1})^{2}.
\end{equation}
Considering the signature convention in NP formalism, we get the mass loss theorem in EMD theories: The mass density at any angle of the system can never increase. It is a constant if and only if there is no news. Our mass-loss formula generalizes the one in \cite{LMW} by removing the constraint of axisymmetry.

As for Maxwell part, we work in retarded radial gauge $A_r=0$. The Maxwell-tensor is constructed as
\begin{multline}
F_{\mu\nu} =(\phi_1 + \xbar \phi_1) (n_\mu l_\nu  - l_\mu  n_\nu )  + (\phi_1 - \xbar \phi_1) (m_\mu \bm_\nu  - \bm_\mu  m_\nu ) + \phi_2 (l_\mu m_\nu  - m_\mu l_\nu)\\ + \xbar\phi_2 ( l_\mu \bm_\nu  - \bm_\mu  l_\nu) + \phi_0 (\bm_\mu n_\nu  - n_\mu \bm_\nu) + \xbar\phi_0 ( m_\mu n_\nu  - n_\mu  m_\nu).\label{m}
\end{multline}
We represent the Newman-Penrose variables of Maxwell parts in terms of the gauge fields $A_\mu$,
\be
\label{ma}
A_u^0=-(\phi_1^0 + \xbar\phi_1^0) ,\;\;\;\; \p_u A_z^0 = - \frac{\phi_2^0}{\bP} ,\;\;\;\;A_z^1=-\frac{\xbar\phi_0^0}{\bP},\;\;\;\;(\p_z A_{\bz}^0 - \p_{\bz} A_z^0)=\frac{ \phi_1^0 - \xbar\phi_1^0}{P \bP }.
\ee
\be
\p_u \left(\frac{A_u^0}{P\bP}\right) = \p_u (\p_z A_{\bz}^0 + \p_{\bz} A_z^0),
\ee
where
\be
A_u=\frac{A_u^0(u,z,\bz)}{r} + \cO(r^{-2}),\;\;\;\;A_z=A_z^0(u,z,\bz) + \frac{A_z^1(u,z,\bz)}{r} + \cO(r^{-2}).
\ee

From eq(\ref{charge}), we find
\begin{equation}
\begin{aligned}{}
\label{cc}
\partial_{u}\phi_{1}^{0}&=\eth\phi_{2}^{0}=P_{s}\partial_{\bz}\phi_{2}^{0}-\partial_{\bz}P_{s}\phi_{2}^{0}\\
&=P_{s}^2\partial_{\bz}(\phi_{2}^{0}/P_{s}).
\end{aligned}
\end{equation}

Taking the real part of eq(\ref{cc}), we get
\begin{equation}
\label{maa}
\partial_{u}(\frac{\phi_{1}^{0}+\xbar\phi_{1}^{0}}{2})=\frac{P_{s}^{2}}{2}(\partial_{\bz}(\phi_{2}^{0}/P_{s})+\partial_{z}(\xbar\phi_{2}^{0}/P_{s})).
\end{equation}
Substituting eq(\ref{ma}) into eq(\ref{maa}), we find
\begin{equation}
\partial_{u}A_{u}^{0}=\partial_{u}(P_{s}^{2}\partial_{\bz}A_{z}^{0}+P_{s}^{2}\partial_{z}A^{0}_{\bz}).
\end{equation}
Defining the flux,
\begin{equation}
\Phi=A_{u}^{0}-P_{s}^{2}(\partial_{\bz}A_{z}^{0}+\partial_{z}A_{\bz}^{0}),
\end{equation}
we find
\begin{equation}
\partial_{u}\Phi=0.
\end{equation}
This means that the flux does not change with time. According to Gauss's Law, we can conclude that the charge is conserved, which again generalizes the result in \cite{LMW} to asymptotically flat case.
\section{The memory effects}
\label{memories}
 According to the solution space in the previous section, we can derive the memory effects. We will examine the motion of a charged time-like particle to specify the observational effects in a unified expression\cite{MT}. That is to say, we consider the contributions of the gravitational radiation, electromagnetic radiation and scalar radiation at the same time by considering the effects of the motion of a charged particle caused by the radiations in EMD theories. Gravitational memory effects, characterized by the non-linear contribution to the overall change in the shear of outgoing null surfaces at the future null infinity\cite{F}, have a large group of observational effects of the gravitational radiation. Displacement memory effects are observational effects about a location displacement of the observers, i.e.\cite{SZ} describes a distance shift of two parallel inertial detectors near the null infinity caused by the radiative energy flux. Spin memory effects\cite{MW,PSZ} are memory effects characterized by the observational phenomena that the radiation causes the observers to rotate. i.e. \cite{PSZ} is about a relative time delay of two beams of light on clockwise and counterclockwise orbits induced by the radiative angular momentum flux. \cite{MW} discovers a kind of spin memory effect characterized by the time delay of a free-falling massive particle constrained on a time-like, $r=r_{0}$ hypersurface. The memory effects also exist in Maxwell theory called electromagnetic memory effects, i.e.\cite{LBDG,MOWW,P,SUS}. \cite{LBDG} is a change of the velocity (a``kick'') of a charged particle. Here we will consider all these three kinds of memory effects. The charged particle will be constrained on a time-like, $r=r_{0}$ hypersurface. $r=r_{0}$ is a fixed radial distance which is very large, means that the particle is very far from the gravitational and electromagnetic source. The induced metric of this hypersurface can be derived by inserting the solution space in the previous section into eq(\ref{eq3}) and eq(\ref{eq111}), which in series expansions is given by
\begin{equation}
\begin{aligned}
\label{metric}
ds^{2}&=[1+\frac{1}{r_{0}}(\Psi_{2}^{0}+\xbar\Psi_{2}^{0}-\frac{1}{3}\varphi_{1}\partial_{u}\varphi_{1})+\frac{1}{12 r^{2}_{0}}(12\phi_{1}^{0}\xbar\phi_{1}^{0}-4\xbar\eth\Psi_{1}^{0}-4\eth\xbar\Psi_{1}^{0}+2\xbar\eth\varphi_{1}\eth\varphi_{1}
\\&+2\varphi_{1}\xbar\eth\eth\varphi_{1}-4\varphi_{2}\partial_{u}\varphi_{1}+2\varphi_{1}\partial_{u}\varphi_{2})+\cO(r_{0}^{-3})]du^{2}+2[-\frac{\eth\xbar\sigma^{0}}{P_{s}}+\frac{1}{6P_{s}r_{0}}(4\xbar\Psi_{1}^{0}-\varphi_{1}\xbar\eth\varphi_{1})\nonumber
\end{aligned}
\end{equation}
\begin{equation}
\begin{aligned}
&+\cO(r_{0}^{-2})]du dz
+2[-\frac{\xbar\eth\sigma^{0}}{P_{s}}+\frac{1}{6P_{s}r_{0}}(4\Psi_{1}^{0}-\varphi_{1}\eth\varphi_{1})+\cO(r_{0}^{-2})]dud\bz+[-2\frac{\xbar\sigma^{0}r_{0}}{P_{s}^{2}}\\
&+\frac{1}{3P_{s}^{2}r_{0}}(\xbar\Psi_{0}^{0}+\xbar\sigma^{0}\varphi_{1}^{2})+\cO(r_{0}^{-2})]dz^{2}+[-2\frac{\sigma^{0}r_{0}}{P_{s}^{2}}+\frac{1}{3P_{s}^{2}r_{0}}(\Psi_{0}^{0}+\sigma^{0}\varphi_{1}^{2})\\
&+\cO(r_{0}^{-2})]d\bz^{2}-2[\frac{r_{0}^{2}}{P_{s}^{2}}+\frac{-\varphi_{1}^{2}+4\sigma^{0}\xbar\sigma^{0}}{4P_{s}^{2}}+\frac{-\varphi_{1}\varphi_{2}}{3P_{s}^{2}r_{0}}+\cO(r_{0}^{-2})]dzd\bz.
\end{aligned}
\end{equation}

One should notice that here we fix the topology of the 2-surface, i.e. $P=\bP=P_{s}=\frac{1+z\bz}{\sqrt{2}}$. The induced Maxwell field on the $r=r_0$ hypersurface is
\begin{equation}
\begin{aligned}
F_{uz}&=-\frac{\phi_{2}^{0}}{P_{s}}+\frac{1}{2P_{s}r_{0}}(a\phi_{2}^{0}\varphi_{1}-2\xbar\sigma^{0}\xbar\phi_{2}^{0}+2\xbar\eth\phi_{1}^{0})+\cO(r_{0}^{-2}),\\
F_{u\bz}&=-\frac{\xbar\phi_{2}^{0}}{P_{s}}+\frac{1}{2P_{s}r_{0}}(a\xbar\phi_{2}^{0}\varphi_{1}-2\sigma^{0}\phi_{2}^{0}+2\eth\xbar\phi_{1}^{0})+\cO(r_{0}^{-2}),\\
F_{z\bz}&=\frac{\phi_{1}^{0}-\xbar\phi_{1}^{0}}{P_{s}^{2}}+\frac{\eth\xbar\phi_{0}^{0}-\xbar\eth\phi_{0}^{0}}{P_{s}^{2}r_{0}}+\cO(r_{0}^{-2}).
\end{aligned}
\end{equation}
The dilaton field on the hypersurface is
\begin{equation}
\varphi(u,r_{0},z,\bz)=\frac{\varphi_{1}(u,z,\bz)}{r_{0}}+\frac{\varphi_{2}(u,z,\bz)}{r_{0}^{2}}+\cO(r_{0}^{-3}),
\end{equation}
where
\begin{equation}
\partial_{u}\varphi_{2}+\eth\xbar\eth\varphi_{1}=0.
\end{equation}
The equation of motion of free falling charged particle on this hypersurface is
\be\label{geodesic}
V^\nu(\xbar\n_\nu V^\mu + q {\xbar F_\nu}^\mu) =0.
\ee
where V is the tangent vector of the particle worldline, $\xbar\n$ is the covariant derivative on this 3 dimensional hypersurface and $q$ is the charge of the particle.

According to \cite{MW}, we impose that $V$ has the following asymptotic expansion
\be
V^u=1 + \sum\limits_{a=1}^\infty\frac{V^u_a}{r^a},\;\;\;\;V^z=\sum\limits_{a=2}^\infty\frac{V^z_a}{r^{a}}.
\ee
Then we solve \eqref{geodesic} order by order. The solution up to relevant order is
\begin{equation}
\begin{aligned}{}
V^{u}_{1}&=-\frac{1}{2}(\Psi_{2}^{0}+\xbar\Psi_{2}^{0})+\frac{1}{6}\varphi_{1}\partial_{u}\varphi_{1},\\
V^{z}_{2}&=-P_{s}\xbar\eth\sigma^{0}+qP_{s}^{2}A^{0}_{\bz},\\
V^{u}_{2}&=q^{2}P_{s}^{2}A^{0}_{z}A^{0}_{\bz}+\frac{1}{6}(\xbar\eth\Psi_{1}^{0}+\eth\xbar\Psi_{1}^{0})+\frac{3}{8}(\Psi_{2}^{0}+\xbar\Psi_{2}^{0})^{2}-\eth\xbar\sigma^{0}\xbar\eth\sigma^{0}-\frac{1}{2}\phi_{1}^{0}\xbar\phi_{1}^{0}\\
&-\frac{1}{4}(\Psi_{2}^{0}+\xbar\Psi_{2}^{0})\varphi_{1}\partial_{u}\varphi_{1}-\frac{1}{12}\varphi_{1}\partial_{u}\varphi_{2}+\frac{1}{6}\varphi_{2}\partial_{u}\varphi_{1}+\frac{1}{24}(\varphi_{1}\partial_{u}\varphi_{1})^{2}\\
&-\frac{1}{12}(\eth\varphi_{1}\xbar\eth\varphi_{1}+\varphi_{1}\eth\xbar\eth\varphi_{1}),\\
V^{z}_{3}&=P_{s}[2\eth\xbar\sigma^{0}\sigma^{0}+\frac{2}{3}\Psi_{1}^{0}+\frac{1}{2}\xbar\eth\sigma^{0}(\Psi_{2}^{0}+\xbar\Psi_{2}^{0})]-P_{s}\int dv\frac{1}{2}(\eth\Psi_{2}^{0}+\eth\xbar\Psi_{2}^{0}+2q\eth A_{u}^{0}),\\
&-2qP_{s}^{2}\sigma^{0}A_{z}^{0}+qP_{s}^{2}A_{\bz}^{1}-\frac{1}{6}P\xbar\eth\sigma^{0}\varphi_{1}\partial_{u}\varphi_{1},
\end{aligned}
\end{equation}
where we have set all integration constants of $u$ to zero since we require that the charged particle is static initially. At $r_0^{-2}$ order, we can see that V has angular components. In other words, gravitational and electromagnetic radiations characterized by $\sigma^{0}$ and $A_{z}^{0}$ cause free falling charged particle to rotate over some tiny angle about the ``center'' of the spacetime $r=0$. The leading memory effect is the velocity kick of the charged particle
\be
\label{lm}
\Delta V^z=-\frac{1}{r_0^2}(P_s \xbar\eth\Delta\sigma^0 - q P_s^2 \Delta A_{\bz}^0) + \cO(r_0^{-3}) .
\ee
 The leading memory effect consists of two parts, namely the gravitational part $-P_s \xbar\eth\Delta\sigma^0$ and electromagnetic part $ q P_s^2 \Delta A_{\bz}^0$. They are mathematically equivalent to leading soft graviton theorem\cite{SZ} and leading soft photon theorem\cite{P} respectively by a Fourier transformation. That is why we call this a unified expression of leading gravitational memory effect and leading electromagnetic memory effect. It is the same as the result in Einstein-Maxwell theory\cite{MT}, which means that the scalar field has no contribution to the leading memory effect. Besides this, we can not see the coupling effect in the leading memory effect. We should notice that the change of the velocity of the charged particle (the velocity kick) is considered as distinct effect from the displacement memory effect. The gravitational memory is a property of gravitational wave characterized by the change of the asymptotic shear $\Delta\sigma^{0}$\cite{F}. The velocity kick we discuss in this paper and the relative displacement of nearby observers(e.g. in \cite{SZ}) are different observational effects of the gravitational wave with memory.

According to the treatment in electromagnetism\cite{MOWW}, the sub-leading memory effect is a position displacement of the charged particle\footnote{We have used the fact that $du=d\chi + \cO(r_0^{-1})$, where $\chi$ is the proper time.}.
\be
\Delta z=\int V^z du=-\frac{1}{r_0^2}\int du (P_s \xbar\eth \sigma^0 - q P_s^2 A_{\bz}^0) + \cO(r_0^{-3}) .
\ee
The gravitational contribution $-\int (P_s \xbar\eth \sigma^0) du$ has a relevance to sub-leading soft graviton theorem(see \cite{PSZ} for specific discussion), and the electromagnetic contribution $\int ( q P_s^2 A_{\bz}^0) du$ has a relevance to subleading soft photon theorem (see \cite{MOWW} for further discussion). So we call this a unified expression of subleading gravitational memory effect and subleading electromagnetic memory effect. The result is the same as\cite{MT}, we do not see the coupling effect in this sub-leading memory effect either.

Another sub-leading observational memory effect  is a time delay of the observe \cite{MW,FGHN}. It is a kind of spin memory effect. The time delay of a charge particle will also have contributions from the electromagnetic radiation and the scalar radiation. Since $V$ is time-like, the infinitesimal change of the proper time can be derived from the co-vector\footnote{We have used the fact that $dz=\frac{V^z_0}{r_0^2}du + \cO(r_0^{-3})$.}
\begin{equation}
\begin{aligned}
\label{spm}
d\chi&=\{1+\frac{1}{6r_{0}}(3\Psi_{2}^{0}+3\xbar\Psi_{2}^{0}-\varphi_{1}\partial_{u}\varphi_{1})+\frac{1}{r_{0}^{2}}[-\frac{1}{8}(\Psi_{2}^{0}+\xbar\Psi_{2}^{0})^{2}-\frac{1}{6}(\xbar\eth\Psi_{1}^{0}+\eth\xbar\Psi_{1}^{0})+\xbar\eth\sigma^{0}\eth\xbar\sigma^{0}\\
&+\frac{1}{2}\phi_{1}^{0}\xbar\phi_{1}^{0}-q^{2}P_{s}^{2}A_{z}^{0}A_{\bz}^{0}+\frac{1}{12}(\eth\varphi_{1}\xbar\eth\varphi_{1}+\varphi_{1}\eth\xbar\eth\varphi_{1})-\frac{1}{6}\varphi_{2}\partial_{u}\varphi_{1}+\frac{1}{12}\varphi_{1}\partial_{u}\varphi_{2}\\
&+\frac{1}{12}(\Psi_{2}^{0}+\xbar\Psi_{2}^{0})\varphi_{1}\partial_{u}\varphi_{1}-\frac{1}{72}(\varphi_{1}\partial_{u}\varphi_{1})^{2}]\}du+\cO(r_{0}^{-3}).
\end{aligned}
\end{equation}
The electromagnetic contribution $ (\frac{1}{2}\phi_1^0\xbar\phi_1^0 - q^2 P^2_s A_z^0 A_{\bz}^0)$ comes one order higher than the gravitational contribution $\frac{1}{2}(\Psi^0_2 + \xbar \Psi^0_2)$ in the $\frac{1}{r_0}$ expansion, but the scalar contribution $-\frac{1}{6}\varphi_{1}\partial_{u}\varphi_{1}$ appears the same order as the gravitational contribution, which means that the scalar effect is stronger than the electromagnetic effect and it is of the same order as the gravitational effect. The coupling constant $a$ does not show in eq(\ref{spm}) which means that we can not find the effect of the non-minimal coupling of the scalar field and the electromagnetic field at this order. We can find a scalar-gravitational coupled term $\frac{1}{12}(\Psi_{2}^{0}+\xbar\Psi_{2}^{0})\varphi_{1}\partial_{u}\varphi_{1}$ which is the same order as the electromagnetic contribution. Except this, we can not find any other term about the coupling of the gravitation and matter fields at this order in this spin memory effect.

\section{Conclusion}
\label{sec:discussion}

In this work, we studied the EMD theory in NP formalism. We derived the NP equations of the EMD theory and obtained the asymptotically-flat solution space. The solution space is an extension of \cite{LMW} in NP formalism.  It allows us to investigate the memory effects in EMD theories. We found that the dilaton did not contribute to the kick memory effect, nor to the displacement memory effect. However, the dilaton contributes to the time-delay memory effect, and the dilaton contribution arises in the same order in the large $r$ expansion as the the gravitational contribution, and it is one order lower than the electromagnetic contribution. Furthermore. we also discovered a scalar-gravitational coupled term in the same order as the electromagnetic contribution in the time-delay memory effect. However, we found that there was no observable effect associated with the non-minimal coupling between the Maxwell and the dilaton field in all of these three memory effects.

\section*{Acknowledgements}
\label{sec:acknowledgements}

\addcontentsline{toc}{section}{Acknowledgments}

The author thanks Pujian Mao for useful discussion and H. Lu for useful suggestions. This work is supported in part by the NSFC (National Natural Science Foundation of China) under the Grant Nos. 11905156 and 11935009.

\bibliography{ref}

\bibliographystyle{utphys}

\end{document}